\documentclass{emulateapj}

\usepackage{times}

\usepackage{color}

\usepackage{graphicx} 
\newcommand{\kms}{\mbox{${\rm km\,s}^{-1}$}}
\newcommand{\doce}{\mbox{$^{12}{\rm CO}$}}
\newcommand{\trece}{\mbox{$^{13}{\rm CO}$}}
\newcommand{\dieciocho}{\mbox{${\rm C}^{18}{\rm O}$}}
\definecolor{Red}{rgb}{0.61,0.12, 0.14}
\definecolor{RoyalBlue}{rgb}{0.25,0.41,0.88}

\slugcomment{}

\shorttitle{HD~142527 cavity}
\shortauthors{Perez et al.}

\begin{document}

\title{CO gas inside the protoplanetary disk cavity in HD~142527: disk
  structure from ALMA}

\author{S. Perez\altaffilmark{1,2}, S.  Casassus\altaffilmark{1,2},
  F. M\'enard\altaffilmark{2,3,4}, P. Roman\altaffilmark{2,6}, G. van
  der Plas\altaffilmark{1,2}, L. Cieza\altaffilmark{2,5},
  C. Pinte\altaffilmark{3,4}, V. Christiaens\altaffilmark{1,2},
  A. S. Hales\altaffilmark{2,7}.}

\affil{$^1$ Departamento de Astronom\'ia, Universidad de Chile, Casilla
  36-D, Santiago, Chile}
\affil{$^2$ Millenium Nucleus ``Protoplanetary Disks in ALMA Early
  Science,'' Universidad de Chile, Casilla 36-D, Santiago, Chile}
\affil{$^3$ UMI-FCA 3386, CNRS/INSU, Casilla 36-D, Santiago, Chile}
\affil{$^4$ Univ. Grenoble Alpes, IPAG, F-38000 Grenoble, France\\
  CNRS, IPAG, F-38000 Grenoble, France}
\affil{$^5$ Universidad Diego Portales, Facultad de Ingenier\'ia, Av. Ej\'ercito 441, Santiago, Chile}
\affil{$^6$ Center of Mathematical Modelling, Universidad de Chile.}
\affil{$^7$ Joint ALMA Observatory, Alonso de C\'ordoba 3107, Vitacura
  763-0355, Santiago, Chile}

\begin{abstract}
  \noindent Inner cavities and annular gaps in circumstellar disks are
  possible signposts of giant planet formation. The young star
  \objectname{HD~142527} hosts a massive protoplanetary disk with a
  large cavity that extends up to 140~au from the central star, as
  seen in continuum images at infrared and millimeter
  wavelengths. Estimates of the survival of gas inside disk cavities
  are needed to discriminate between clearing scenarios. We present a
  spatially and spectrally resolved carbon monoxide isotopologue
  observations of the gas-rich disk \objectname{HD~142527}, in the
  $J=2-1$ line of \doce, \trece\ and \dieciocho, obtained with the
  Atacama Large Millimeter/submillimeter Array (ALMA).  We detect
  emission coming from inside the dust-depleted cavity in all three
  isotopologues.  Based on our analysis of the gas in the dust cavity,
  the \doce\ emission is optically thick, while \trece\ and
  \dieciocho\ emission are both optically thin. The total mass of
  residual gas inside the cavity is $\sim$1.5-2 M$_{\rm Jup}$.  We
  model the gas with an axisymmetric disk model.  Our best fit model
  shows that the cavity radius is much smaller in CO than it is in
  millimeter continuum and scattered light observations, with a gas
  cavity that does not extend beyond 105~au (at 3-$\sigma$).  The gap
  wall at its outer edge is diffuse and smooth in the gas
  distribution, while in dust continuum it is manifestly sharper.  The
  inclination angle, as estimated from the high velocity channel maps,
  is 28$\pm$0.5 degrees, higher than in previous estimates, assuming a
  fix central star mass of 2.2~M$_\odot$.
\end{abstract}


\section{Introduction}\label{sec:intro}

Planets are thought to form within circumstellar material, in
so-called protoplanetary disks \citep[e.g.][]{Arm2010}. As these
planets grow massive, they imprint a series of morphological and
dynamical features onto their parent disk. During their formation
epoch, one or multiple massive planets could open a gap in a gas-rich
protoplanetary disk \citep{Var2006, Zhu2011}.  The surface density
profile of the gap is determined by the balance between gravity,
viscous and pressure torques \citep{Cri2006}. Evidence of these gaps
has been observed at various wavelengths in nearby young stars
\citep[e.g., ][]{And2011}. 

The star HD~142527, with spectral type F6IIIe, hosts a protoplanetary
disk with a large infrared excess \citep{Mal1999,Fuk2006}. It is a
young gas-rich system, with an estimated age of $\sim$2\,Myr
\citep{Fuk2006}.  HD~142527 is most likely embedded inside the Lu22
condensation of the Lupus dark cloud, at a distance of 140~pc away,
which has a reported $V_{\rm LSR}=4.69 \pm 0.7~\kms$
\citep{Vil2000}. Its disk is almost face-on with an inclination of
$\sim$20$^{\circ}$, from previous infrared data. The disk exhibits a
remarkably large dust-depleted cavity extending out to a radius of
$\sim$140~au, as seen in IR imaging \citep{Fuk2006}, sub-millimeter
continuum and CO line emission \citep{Oha2008, Cas2013}. The outer
disk has a complex non-axisymmetric surface density showing several
spiral arms and clumps \citep{Ram2012, Ave2014}, signs of dynamical
perturbation possibly caused by planetary-mass bodies inside the
cavity or gravitational instability \citep{Chr2014}.

Recent ALMA observations of HD~142527 have evidenced what could be key
stages of the planet formation process: the discovery of gaseous flows
inside the dust-depleted cavity, as seen in HCO+ emission by
\citet{Cas2013}. Resolved CO~$J=3-2$ observations show a peak in
emission in the cavity, in agreement with on-going dynamical clearing
by planetary-mass bodies \citep{Bru2013}. Also using ALMA,
\citet{Fuk2013} found that the $J=3-2$ isotopologues, $^{13}$CO and
C$^{18}$O, are optically thick in the outer disk.

Detection of gas inside a protoplanetary gap provides quantitative
information on the total gas mass content and temperature profile,
depending on whether the line emission is optically thin or optically
thick \citep{Wil2009}. Direct images and estimates of the survival of
the gas inside transition disk cavities are needed to discriminate
between scenarios of dust removal, such as grain growth,
photo-evaporation and planet-disk interactions. For instance, grain
growth is not expected to markedly affect the gas density profile,
while photoevaporation removes both dust and gas \citep{Bru2013}. One
way of discriminating between a planet-induced gap and a
photoevaporation-induced gap can be based on the sharpness of the
transition in surface density at the outer edge of the cavity:
dynamical clearing would result in a rather sharp transition if the
putative planet is massive \citep{Pas2011}.  Millimeter observations
of the rotational gas lines are ideal to quantify the amount of gas
inside a dust-depleted cavity. Moreover, by comparing isotopologue
emission from specific regions of the disk in protoplanetary systems
we can constrain physical properties of the disk, such as its gas
density and gas temperature.

Resolved observations of optically-thin emission can directly trace
the underlying mass density distribution in the disk cavity. The
amount of gas inside a protoplanetary cavity can provide constrains on
planet formation and migration scenarios.  For example, a planet that
is not massive enough to cleanly evacuate the material inside the gap
will only decrease the disk surface density.  Also, such planet is
expected to co-orbit with a corotation zone, which should carry
significant amounts of gas \citep{Arm2011}. Thus, studying the gas
surface density structure of transition disks is of direct importance
to distinguish between different clearing and formation scenarios.

In this paper, we analyse ALMA Band-6 ($\sim$230~GHz) CO line emission
and dust continuum observations of HD~142527. The data provide
resolved observations of gas inside a dust depleted protoplanetary
disk cavity.  Our analysis aims to measure the amount of gas in this
cavity and provide a more complete picture of the structure of the
HD~142527 disk. The interferometric ALMA data reduction and
calibration steps are described in Section~\ref{sec:data}. A detailed
description of the CO emission maps, the morphology and kinematics of
the observed gas, along with diagnostics of the physical conditions
inside the cavity are provided in Section~\ref{sec:results}.  We
develop a disk structure model consistent with HD~142527 observations
to study the properties of the disk. This model and the radiative
transfer calculations are presented in Section~\ref{sec:model}. Data
and modelling are combined and interpreted in the context of gas
depletion mechanisms in Section~\ref{sec:discussion}.
Section~\ref{sec:summary} summarizes the main results of this
investigation. A brief description of HD~142527's spectral energy
distribution, as computed for the model presented in
Section~\ref{sec:model}, is presented in Appendix A.

\section{ALMA data and processsing}\label{sec:data}

\begin{figure*}
\epsscale{0.9}
\plotone{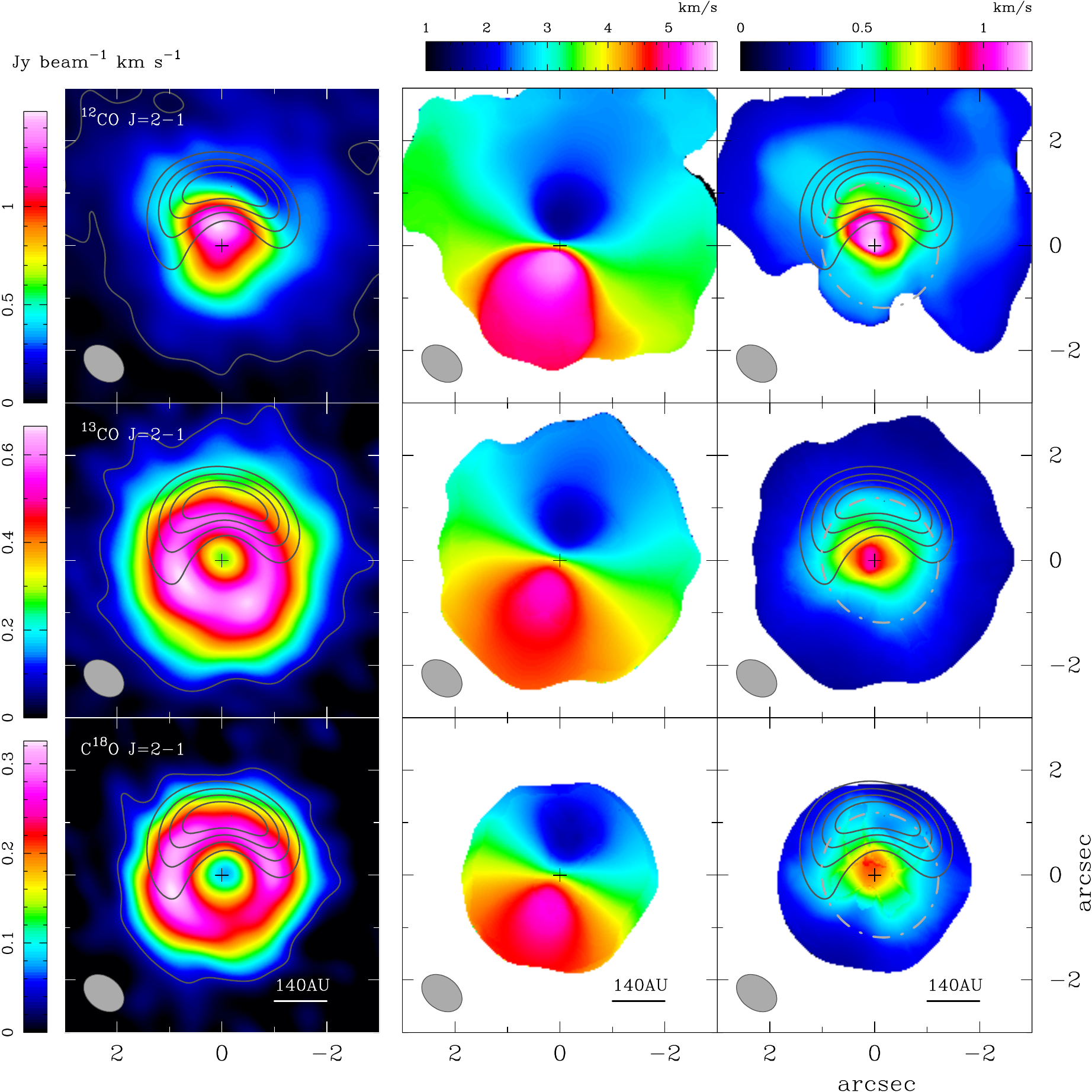}
\caption{Moment maps carbon monoxide isotopologues \doce, \trece\ and
  \dieciocho~$J=2-1$.  North is up, East is left.  {\it Left:} Moment
  zero; continuum subtracted integrated line emission, considering
  flux contribuition from all channels from -0.8 to +7.8~\kms, in
  units of Jy~beam$^{-1}$ \kms. Continuum at 230~GHz is shown in
  contours. The \doce\ map shows the large extent of the molecular
  line emission, the north-south asymmetry is due to foreground
  absorption. \trece\ and \dieciocho\ show a central cavity. The noise
  level for all intensity maps is about 1$\sigma = 11$~mJy per
  beam. {\it Center:} First moment showing the velocity map. {\it
    Right:} Second moment, showing the velocity dispersion of the
  emitting gas. Colour scale is linear. The coordinates origin is set
  to the center of the disk and it is marked with a cross. The
  synthesized beam is shown in the lower left corner. The dashed
  ellipse in the moment 2 map is a fit by inspection of the
  dust-continuum horseshoe border. The ellipse shows that there is a
  difference in dispersion of the gas under the horseshoe, with
  respect to the south counterpart of the disk.}\label{fig:moments}
\end{figure*}

The HD~142527 band 6 observations were acquired on 2012 April 12, and
June 15, using sixteen and twenty 12m antennas respectively. In both
observing sessions the precipitable water vapor in the atmosphere was
stable between 1.5 and 2.0~mm with clear sky conditions, resulting in
median system temperatures ranging from 70 to 90~K. The ALMA
correlator was configured in the Frequency Division Mode (FDM) to
provide 125~MHz bandwidth in four different spectral windows at
0.2~km~s$^{-1}$ resolution per channel. Three spectral windows were
positioned in order to target the $^{12}$CO~$J=2-1$, $^{13}$CO~$J=2-1$
and C$^{18}$O~$J=2-1$ isotopologues at 230.538, 220.399 and
219.560~GHz, respectively.  In both sessions Titan was observed as
flux calibrator, while the quasars 3C279 and J1604-446 were observed
for bandpass and phase calibration respectively. Observations of the
phase calibrator were alternated with the science target every 8
minutes to calibrate the time dependence variations of the complex
gains. The total time spent on-source was 82 minutes.  However, the
concatenated dataset shows spurious spectral line features. Therefore,
in this work we only used the June observations which have higher
signal-to-noise and better $uv$-coverage than the April data.

All the line data were calibrated using the {\it{Common Astronomy
    Software Applications }} package \citep[CASA{\footnote{\tt
      http://casa.nrao.edu/}};][]{2007ASPC..376..127M} in a standard
fashion, which included offline Water Vapor Radiometer (WVR)
calibration, system temperature correction, as well as bandpass, phase
and amplitude calibrations. The fluxes derived for 3C279 in April and
June were 15.9 and 17.5~Jy respectively. For J1604-446, the flux
values after bootstrapping to the observations of Titan were 0.56 and
0.6~Jy in April and June.  By considering the scatter in the fluxes
derived for the bandpass and phase calibrators in the different
observing sessions, we estimate the absolute flux calibration to be
accurate within $\sim$10~$\%$.

Imaging of the CO lines was performed using the CLEAN task in CASA
\citep{1974A&AS...15..417H}. The dataset provides baselines up to 402
meters, which resulted in a synthetic beam size of
0.85$\arcsec\times$0.64$\arcsec$ at PA$=-66.5$~degrees. Continuum
subtraction in the visibility domain was performed prior to imaging of
the CO lines. After CLEANing the images, an RMS noise level of 11~mJy
per 0.2~km~s$^{-1}$ channel was reached.  In this paper, we also make
use of the 345~GHz continuum data described in \citet{Cas2013}.

The central cavity is resolved in the isotopologues data, with
approximately four beams covering the gap. An astrometric correction
was applied to the data set in order to account for the star's proper
motion of $(-11.19, -24.46)$ mas~yr$^{-1}$, as measured by Hipparcos
\citep{vLee2007}. This implied a shift in the images of
0.3~arc-seconds since epoch J2000.

\begin{figure}
\epsscale{1}
\plotone{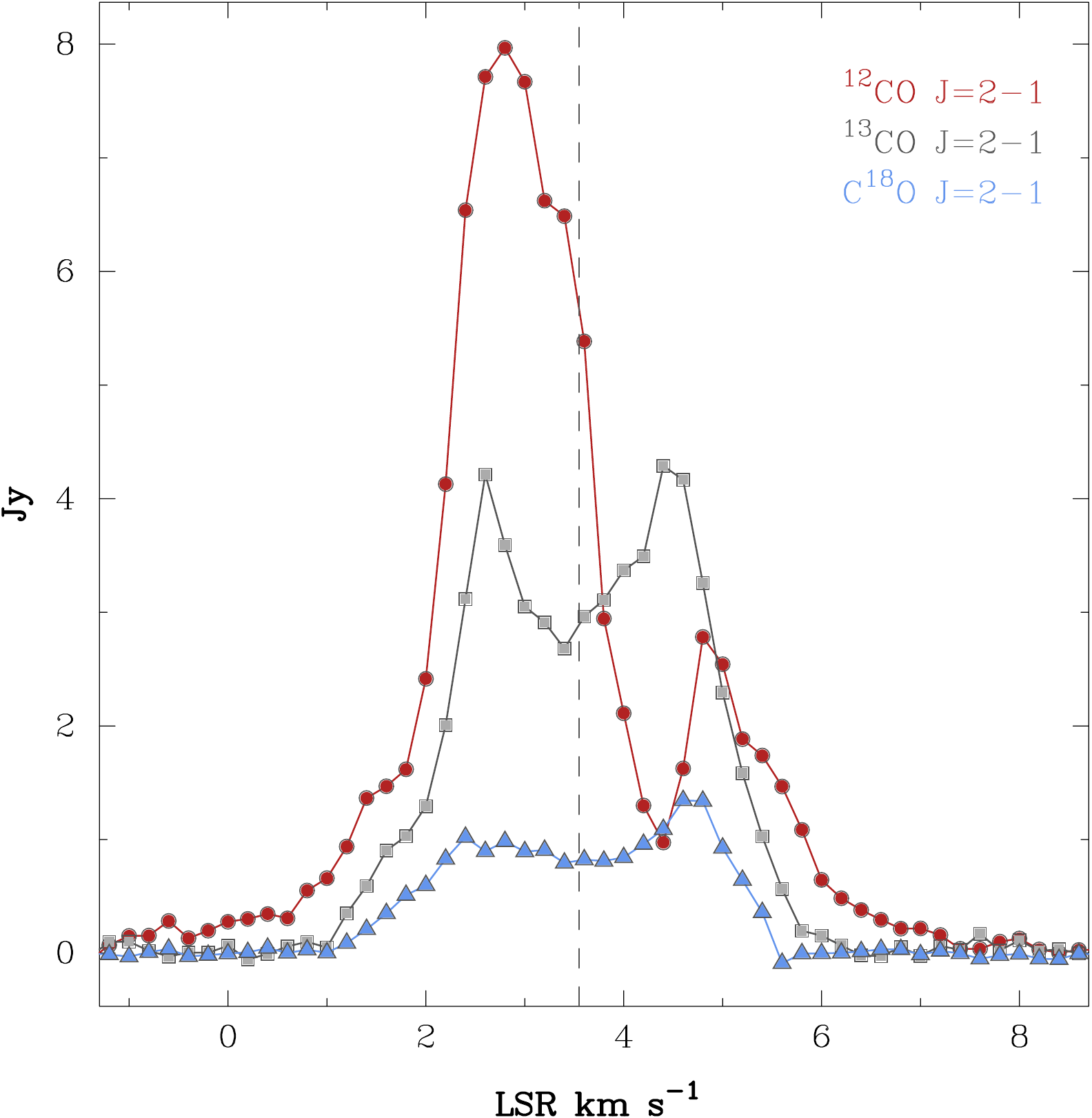}
\caption{ Continuum-subtracted \doce\ (circles), \trece\ (squares) and
  \dieciocho\ (triangles) $J=2-1$ emission line spectra. The spectra
  are extracted summing over the whole data cube. Vertical dotted line
  corresponds to systemic velocity of 3.6~\kms, as determined
  from the \doce\ channel maps.}\label{fig:specs}
\end{figure}

\section{Results}\label{sec:results}

\begin{deluxetable}{lccccc}
\tabletypesize{\scriptsize}
\tablecaption{Properties of emission line and continuum maps}
\tablewidth{0pt}
\tablehead{
\colhead{Line} & \colhead{$\nu$} & Beam  & \colhead{$1\sigma$ Noise} & 
\colhead{Peak $I_\nu$} & Flux$^{\rm a}$ \\
$J$=2--1 & GHz & '' & Jy beam$^{-1}$ & Jy beam$^{-1}$ & Jy}
\startdata

$^{12}$CO & 230.538 & 0.85$\times$0.64 & 0.012 & 0.74 & 17.4\\
$^{13}$CO & 220.399 & 0.85$\times$0.64 & 0.013 & 0.70 & 11.4\\
C$^{18}$O & 219.560 & 0.85$\times$0.64 & 0.013 & 0.39 & 3.4\\
\hline
\multicolumn{6}{c}{Continuum}\\
\hline
Cont. & 225.393 & 0.94$\times$0.72 & 0.0003 & 0.27 & 1.09 
\enddata
\tablecomments{ Line emission and continuum maps properties. Maps were
  cleaned with CASA and binned to $\Delta v = 0.20$~km~s$^{-1}$
  spectral resolution. $^{\rm a}$ Total flux integrated over the whole
  data cube ($\sim$10''x10'' aperture).}
\label{tab:maps}
\end{deluxetable}

\subsection{Morphology and Kinematics of the Gas Disk}

Molecular gas observations, such as millimeter/submillimeter pure
rotational lines of CO, can provide constrains on the amount of gas
mass inside transition dust-depleted disk cavities.  The dust-depleted
cavity in HD~142527 has an apparent size of $\sim$2'' across
\citep{Cas2012}. Our ALMA Cycle 0 observations have enough sensitivity
and resolution to detect and resolve such left-over gas in HD~142527
cavity.

\subsubsection{Moment maps}

Fig.~\ref{fig:moments} shows the integrated intensity, velocity
centroid, and velocity dispersion for the three isotopologues. All the
moment maps velocity information displayed in Fig.~\ref{fig:moments}
were derived by fitting a gaussian profile to the spectral axis in
each spatial pixel of the CLEANed data cubes.  The maps were then
median-filtered with a kernel box of 0.1$\times$0.1''.  We detect
extended \doce\ emission, 5-$\sigma$ above the noise level, as far as
5'' away from the center, which corresponds to $\sim$700~au
\citep[see][for a discussion of the extended features seen in
  \doce]{Chr2014}.

The \doce~$J=2-1$ integrated emission (Fig.~\ref{fig:moments}, or
zeroth-order moment map $I(\doce)$ (Fig.~1, upper left) is smooth,
with no counterpart of the cavity seen in the CO isotopologues. In
fact, $I(\doce)$ peaks inside the dust-depleted cavity, at a location
close to the star but intriguingly shifted to the north by 0.4''
(about half a beam). The stellar position, as tabulated in the SIMBAD
database, is consistent with the disk's expected center as given by
the center of symmetry in the velocity field (represented with a black
cross in Fig.~\ref{fig:moments}).  The lack of a decrement in
$I(\doce)$ outlining the cavity, as seen in its rarer isotopologues,
suggests that the line is most likely optically thick (See
Section~\ref{sec:thick}).

The asymmetric morphology of the emission and its off-centered peak
are due to absorption by an intervening cloud, at
4.4~\kms\ \citep{Cas2013Champ}, thus affecting the red part of the
line.  The foreground cloud can be seen as an absorption feature in
the \doce\ spectrum shown in Fig.~\ref{fig:specs}.  We measure the
systemic velocity at $3.6\pm0.1$~\kms\ from the integrated CO spectra,
by assuming that the line wings are symmetric (see
Fig.~\ref{fig:specs}). The error is given by half the size of a
spectral bin.

The HD~142527 dust continuum at 230~GHz, shown as contours in
Fig.~\ref{fig:moments}, has a horseshoe-shaped morphology, as
previously reported in \citet{Cas2013} at 345~GHz. The contours
delineate the dust-depleted cavity, with a radius of $\sim$1'' and a
contrast of $\sim$25 between the north-eastern maximum and the
south-western minimum, slightly shallower than the contrast of 30
reported at 345~GHz \citep{Cas2013, Fuk2013}. The \trece\ integrated
intensity map (Fig.~\ref{fig:moments}) shows a disk cavity and a
bright outer disk. The outer disk is at least a factor of 2 brighter
than the inner cavity in \trece. This is a lower limit since the gap
edge is naturally convoluted with the CLEAN beam, smoothing out the
sharpness of the gap wall.

The velocity dispersion map of \doce\ shows an increment in the width
of the emission line profile under the horseshoe-shaped continuum (see
dashed ellipse in Fig.~\ref{fig:moments} upper right). This wider
emission line can also be seen in the \trece\ dispersion map (middle
right panel), which is less affected by foreground absorption.

\subsubsection{Channel maps}
\label{sec:channels}
\begin{figure*}
\epsscale{1.1} 
\plotone{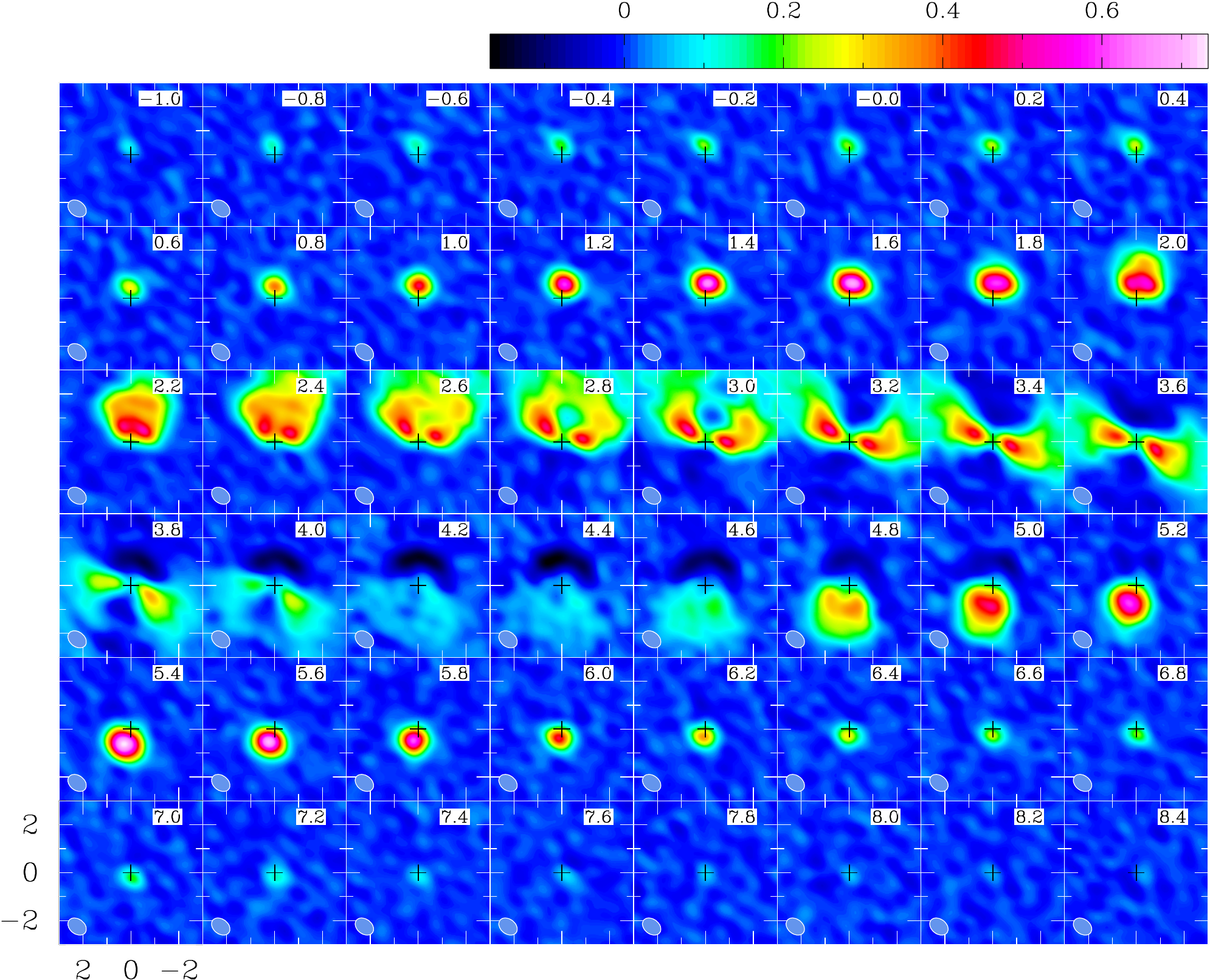}
\caption{\doce~$J=2-1$ channel maps from CLEAN reconstruction in an
  exponential color scale. The LSR velocity for each channel is
  indicated in the upper right-hand corner in \kms. North is up, East
  is to the left. The systemic velocity of HD~142527 is
  3.6~km\,s$^{-1}$. The synthesized beam is shown in the lower
  left-hand corner. The cross at origin represents the stellar
  position which most likely represents the center of rotation of the
  disk velocity profile. The color wedge is in units of
  Jy~beam$^{-1}$~\kms. The position of the disk center, which we
  assume is given by the center of symmetry of the emission at
  systemic velocity (channel at +3.6~\kms), is consistent with the
  estimated astrometry after correcting by the source's proper motion
  (black cross).  \label{fig:map:12CO}}
\end{figure*}

\begin{figure*}
\epsscale{1.1}
\plotone{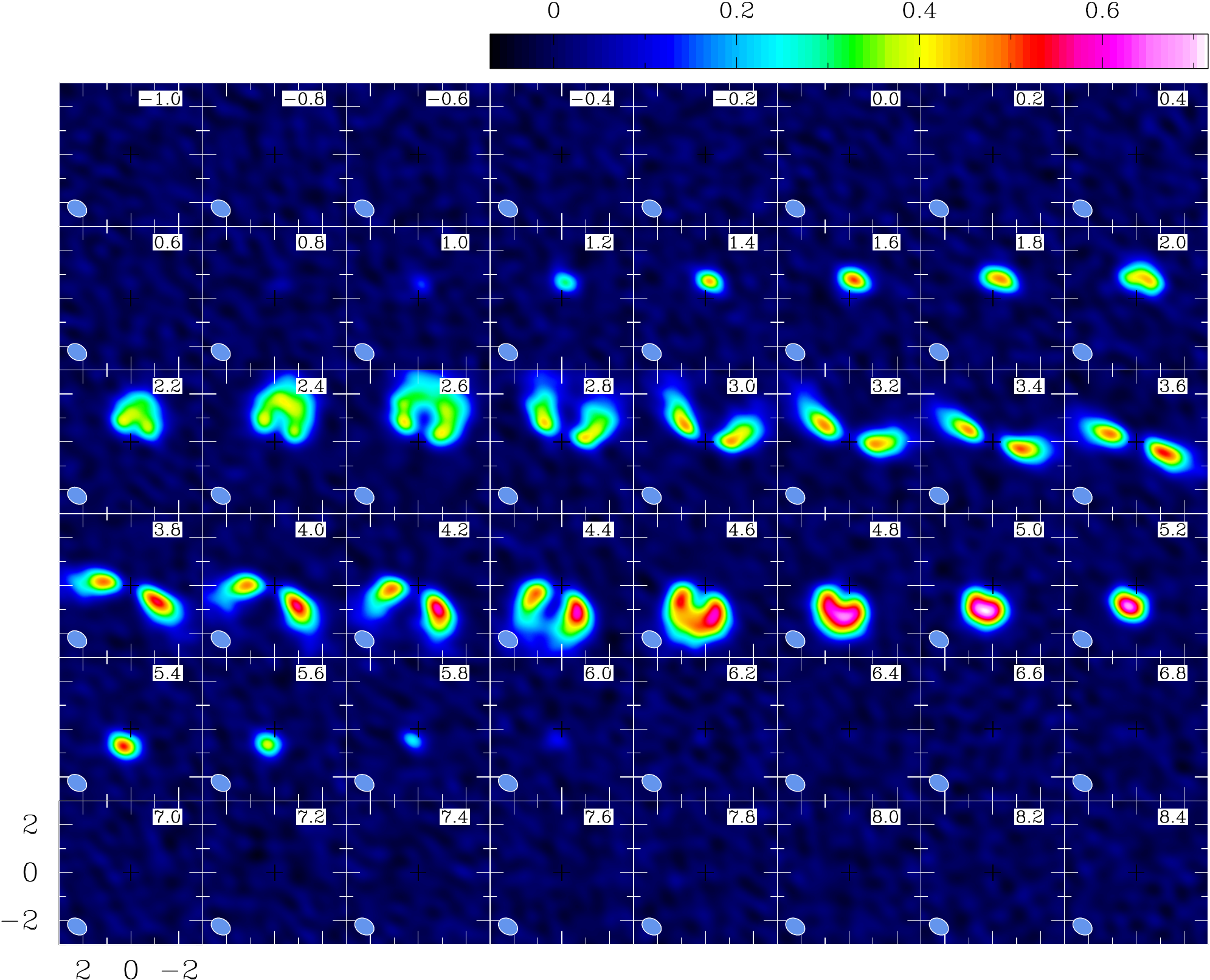}
\caption{\trece\ channel maps in exponential color scale. For details,
  see caption in Fig.~\ref{fig:map:12CO}.\label{fig:map:13CO}}
\end{figure*}

\begin{figure*}
\epsscale{1.1}
\plotone{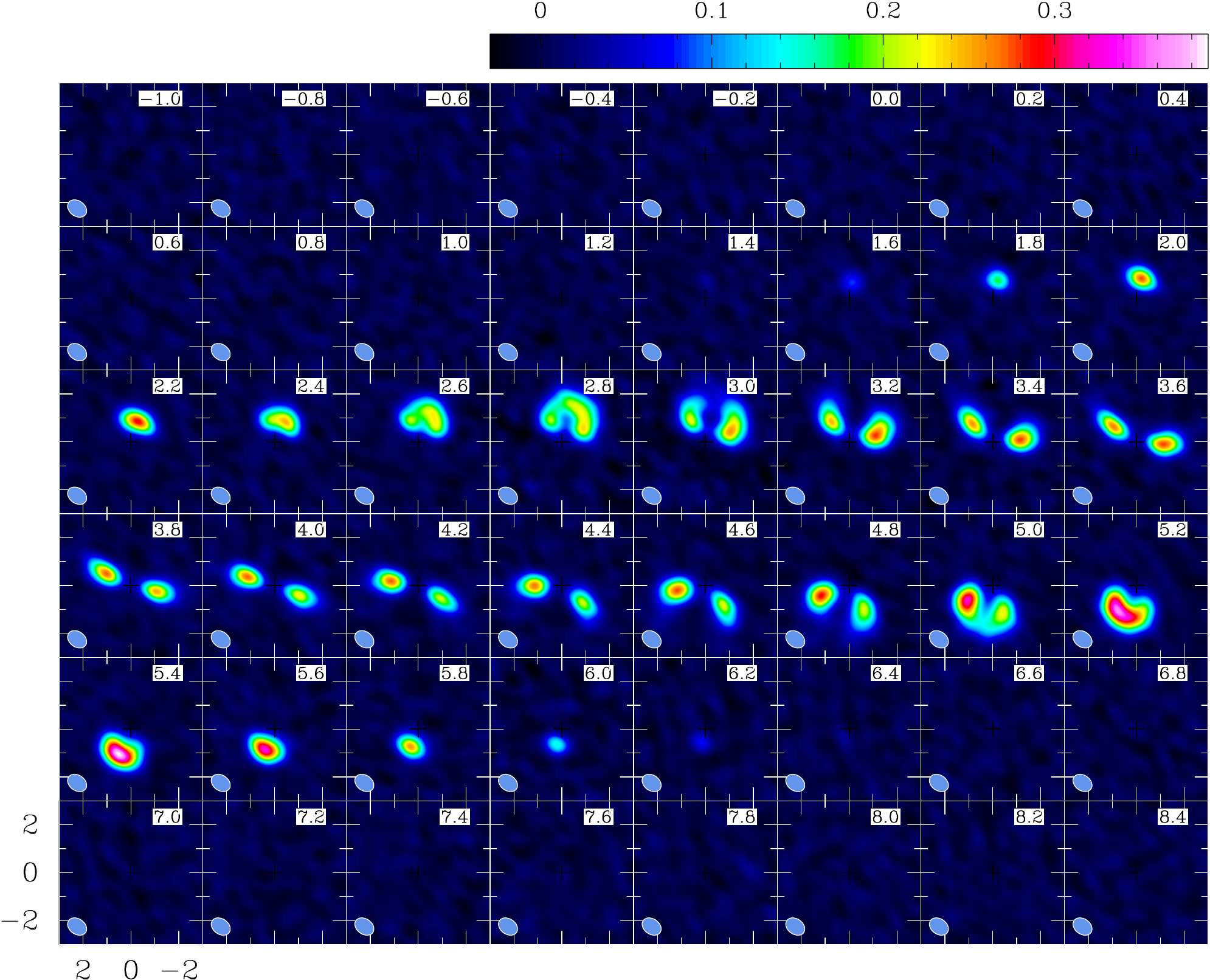}
\caption{\dieciocho\ channel maps in linear color scale. For details,
  see caption in Fig.~\ref{fig:map:12CO}.\label{fig:map:C18O}}
\end{figure*}

Figures~\ref{fig:map:12CO}, \ref{fig:map:13CO} and \ref{fig:map:C18O},
show channel maps for \doce, \trece\ and \dieciocho\ emission from
HD~142527. The Keplerian disk can be identified in the channel maps
as butterfly-shaped emission.  Properties of the maps, rms noise
estimates, peak and integrated line intensities for each isotopologue,
are listed on Table~\ref{tab:maps}. Data were binned to 0.2~\kms\ per
channel in LSR velocity. Maps were reconstructed with CLEAN using
Briggs weighting. The CO lines were detected at 3$\sigma$ level over a
broad velocity range. \doce\ was detected from -0.8 to +7.8~\kms\ in
LSR. \trece\ was detected between +1.0 and +6.2~\kms, while
\dieciocho\ between +1.0 and +5.4~\kms.

An emission line originating from a disk in Keplerian rotation is
expected to show a dipole field pattern on the surface of the disk,
when binned in velocity \citep{Hor1986}. The low velocity channels,
close to systemic velocity $\sim$3.6~\kms, have a Keplerian rotation
profile, along a NE to SW position angle. Gas emission coming from
inside the 140~au dust-depleted cavity can be identified in the high
velocity channel maps. Assuming a Keplerian velocity profile and the
system parameters listed in Table~\ref{tab:params}, the observed
emission from channels at velocities $<+1.8$~km~s$^{-1}$ and
$>+5.4$~km~s$^{-1}$, is radiated by gas material contained in a radius
of 140 au ($\sim 1$''), thus coming entirely from inside the
dust-depleted cavity.

We note that gas emission coming from very near the central star shows
signatures of non-Keplerian kinematics.  This is evidenced by a change
in position angle of the emission in the high velocity channels.  For
instance, the +0.0~\kms\ channel map has a PA that is shifted towards
the NE, when comparing to the systemic velocity PA (near channel
+3.6~\kms). Moreover, the red counterpart of that high velocity
channel, the +7.2~\kms channel, has a PA shifted in the opposite
direction, toward SW. If the gas inside the cavity is indeed being
accreted onto the star, the infalling material would have a radial
velocity component that may explain the non-Keplerian emission at high
velocity channels \citep{Ros2014}.

Another evident feature of the gaseous disk in HD~142527 can be seen
in the \trece\ and \dieciocho\ channel maps.  There is a decrement,
meaning lower emission at a certain blue velocity channel when
compared to its red counterpart, towards the NE of the disk. Compare,
for example, channel +2.4 with +4.8~\kms\ in both isotopologues, where
the red channel is a factor of nearly 2 brighter than the blue (also
seen as a slight asymmetry in the integrated line profiles in
Fig~\ref{fig:specs}). This decrement is, to a moderate extent,
coincident with the horseshoe-shaped continuum.  Furthermore,
\citet{VdP2014} presented HD~142527 imaging of the HCN~$J=4-3$ and
CS~$J=7-6$ emission lines and showed that these lines are also
suppressed under the horseshoe-shaped continuum emission peak.

\subsection{A cut through the inner and outer disks}
\label{sec:cut}

\begin{figure*}
\epsscale{0.8}
\hspace*{-2mm}
\plotone{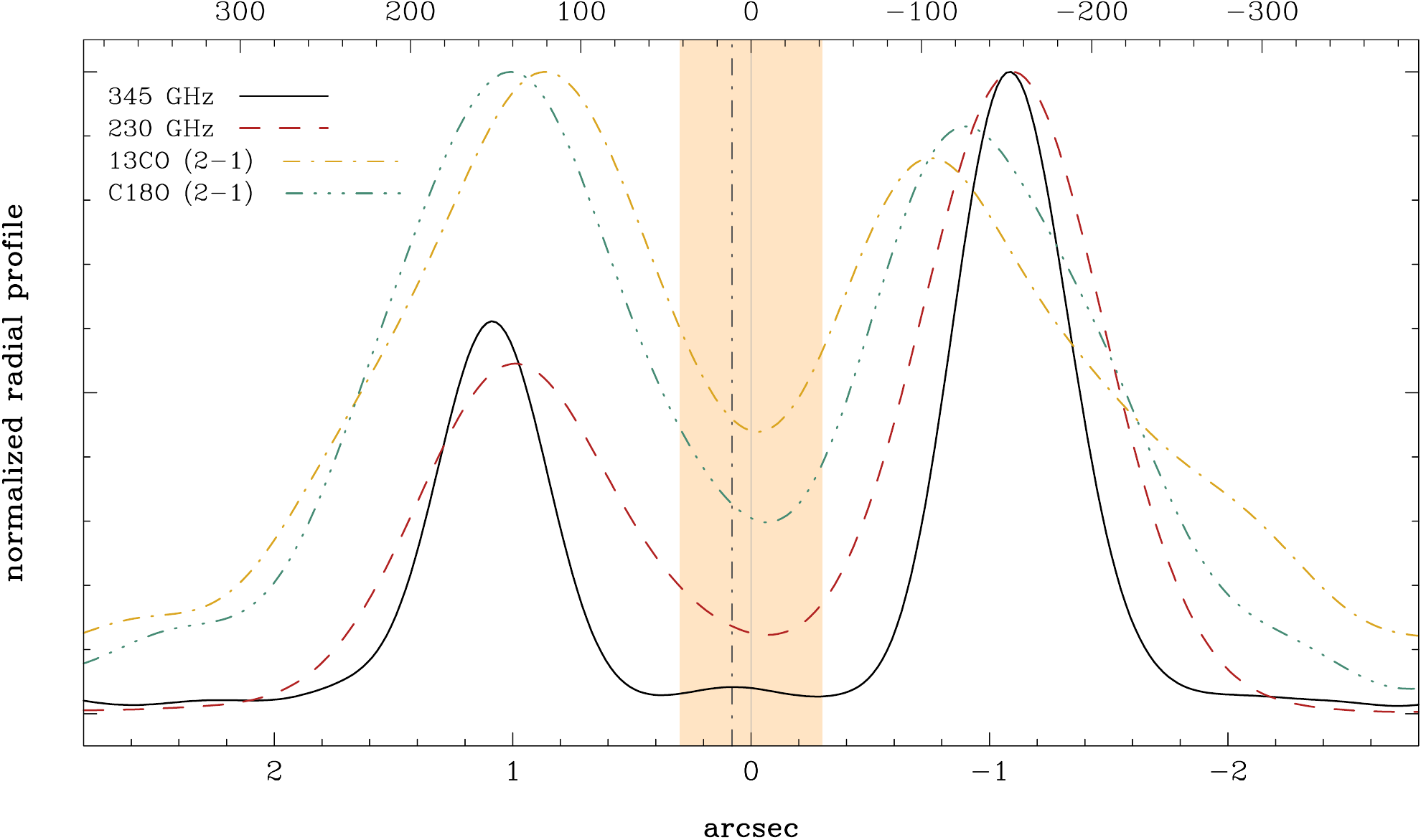}
\caption{HD~142527 disk profile cuts extracted from the zeroth order
  moment maps (at 2h). Solid black and dashed red lines show
  integrated intensity of 345~GH and 230~GHz continuum,
  respectively. \trece\ and \dieciocho\ $J=2-1$ emission integrated
  intensity cuts are shown in yellow dot-dash and green dot-dot-dash
  lines. The shaded region shows the size of the beam across the cut
  axis. The extracted cuts have been scaled to the 345~GHz continuum
  peak to aid comparison between gas emission and dust. The bottom and
  top $x$-axes are in arcsec and au, respectively. The beam size at
  345~GHz is 0.51''$\times$0.33''. The $x$-axis origin (solid vertical
  line) is given by the mid point in between the two 345~GHz peaks,
  determined by fitting single gaussians. The vertical dot-dashed line
  indicates the centroid of the faint central emission in 345~GHz
  continuum (see Section~\ref{sec:cut}).}
\label{fig:cut}
\end{figure*}

Figure~\ref{fig:cut} shows a cut through the disk (near the major
axis, at 2h) in integrated intensity.  The profiles correspond to dust
continuum at 345 and 230~GHz (black solid and dashed red lines,
respectively), and, \trece\ and \dieciocho\ isotopologues
emission. All the profiles are double-peaked, with the peaks
asymmetrical with respect to the center of the disk. The center of the
disk is assumed to be the mid point in between the two 345~GHz peaks.
The dust emission is 40\% weaker in the SE (left peak) than the NW
(right peak), in opposition to the line emission which appears 10-15\%
stronger in the SE. The \doce\ profile peaks inside the cavity, but it
is heavily affected by foreground absorption, hence it is not included
in Figure~\ref{fig:cut}.

The 345~GHz (870~$\mu$m) continuum cut shows an unresolved compact
emission near the center of the cavity.  The centroid of this emission
is shifted from the apparent center of the outer disk as given by the
mid point between the bright outer disk peaks (see vertical dot-dash
line in Fig.~\ref{fig:cut}). The mid point between the peaks also
coincides with the center of rotation of the CO gas. The shift is
$\sim$0.1'' to the East and corresponds to a deprojected distance of
$\sim$16~au at 140~pc.


We roughly quantify the position and roundness of the disk wall in the
continuum and in line emission by fitting a simple gaussian function
to each profile.  The full-width at half-maximum (FWHM) of the 230~GHz
continuum peaks are both about 0.9'', so just resolved by our beam
(the cut through the disk is along the beam's semi-minor axis), with
peak centroids at -1.07'' and 1''. The 345~GHz continuum (band 7) peak
widths are 0.57'' and 0.53'', for the West and East peaks,
respectively.  

The double-peaked morphology in \trece\ and \dieciocho\ in
Fig.~\ref{fig:cut} reveals a smaller and shallower gaseous cavity than
the dust-depleted cavity traced by the continuum.  Interestingly, the
line emission presents a more complex profile than the continuum which
we accounted for by fitting a gaussian on top of a variable order
polynomial function. The FWHM of the peaks in \trece\ are $\sim$1.2''
centered at -0.75'' and 0.9'' away from the disk origin, while
\dieciocho\ is slightly narrower with FWHM of $\sim$1.1'' centered at
-0.9'' and 1'', closer to the disk center.  The CLEANed beam for the
CO isotopologues (shaded region shown in Fig.~\ref{fig:cut}) is only
0.6'' across the cut axis, suggesting that the gap profile is indeed
resolved by our gas observations.  Therefore, the gas outer disk inner
wall lies closer to the star, at 90~au.  The CO flux density appears
more tapered towards the centre than the millimeter dust. And, the
outer disk inner wall is significantly broader in gas, 1.2'' in
\trece, than in millimeter continuum, 0.9'' at 230~GHz.

\vfill

\vfill

\clearpage

\subsection{Optical Thickness of the Lines}\label{sec:thick}

We can estimate the optical depths ($\tau_\nu$) of each isotopologue
from their observed isotopic line ratio \citep[e.g.,][]{Lyo2011}. For
this, we assume CO gas is distributed as an isothermal slab that is
larger than the telescope beam.  In this case, the isotopic line ratio
$R$ can be written as
\begin{equation} 
  R = \frac{T_{\rm B}(\nu_1)}{T_{\rm B}(\nu_2)} = \left (
  \frac{\nu_1}{\nu_2}\right ) \frac{1-e^{-\tau_1}}{1-e^{-\tau_2}},
\end{equation}

\noindent where sub-indices '1' and '2' refer to two of the
isotopologues, either \doce\ and \trece, or \trece\ and \dieciocho,
respectively. We also assume that the two selected lines share the
same excitation temperature, and that this temperature is high enough
to neglect the contribution from the microwave background
radiation. In this work, we adopt the \doce\ to \trece\ ratio of
$76.27\pm1.94$ as measured in the ISM by \citet{Sta2008}. The
\doce\ to \dieciocho\ ratio is assumed to be the canonical value of
$500$ for the ISM \citep{Wil1994}.

The emission from channels at high rotational velocities, specifically
at velocities $<+1.8$~km~s$^{-1}$ and $>+5.4$~km~s$^{-1}$, correspond
to gas emitted solely from inside the dust cavity. From these
channels, we can estimate the line ratios for the residual gas.  In
this velocity range, the isotopic line ratio $R$ has a value of
2.0$\pm$0.2, which yields optical depths of $\tau_{\scriptstyle \rm
  {12}CO} \sim 56$ and $\tau_{\scriptstyle \rm {13}CO} \sim 0.7$.
Therefore, we conclude that inside the disk cavity \doce\ is optically
thick, while \trece\ is mostly optically thin.

\subsubsection{Temperature profile from $^{12}$CO}
\label{sec:temp}

The observed $^{12}$CO line is optically thick, hence the observed
intensity map of the line traces the gas surface temperature rather
than the underlying density distribution. The temperature profile for
\doce\ seems to be nearly flat, $T(r)\sim \rm{constant}$, well inside
the cavity (as measured from the emission in the high velocity
channels), while it decreases with radius as $r^{-0.3}$ over the disk
inner wall and outer disk, as estimated from the channel map at
systemic velocity.  The peak flux density of $\sim$1 Jy implies a gas
temperature of 43~K for the \doce\ gas surface, measured at a radius
of 80~au.

\subsection{Gas Mass Inside the Gap from $^{13}$CO}

The observed $^{13}$CO line is optically thin inside the disk cavity,
hence it gives an estimate of the gas mass content. The integrated
line intensity of \trece\ emission, $\hat{I} = \sum_v I(v) \Delta v$,
where $I(v)$ is the specific intensity measured at a channel with
velocity $v$, and it is directly proportional to the total column
density of the \trece\ gas. The channel width is $\Delta v =
0.2$~\kms. Assuming the CO gas is near LTE, the column density at the
lower level $J=1$ of the \trece\ molecule can be expressed by
\begin{equation}
  \label{eq:N_l}
  N(1) = \frac{4 \pi \, \hat{I}}{h \nu A_{21}},
\end{equation}
\noindent where $N(1)$ is the column density in level 1 expressed in
m$^{-2}$ units, $A_{21}$ is the line strength given by the Einstein
coefficient for spontaneous emission. The units for $\nu$ are
Hz. Assuming that all energy levels are populated under LTE, the total
population is given by the measured column density for a particular
state J (in our case $J=1$),
\begin{equation}
N({\rm total}) = N(J) \frac{Z}{2J + 1} \exp{\left[ \frac{hB_e
      J(J+1)}{kT} \right]},
\end{equation}
\noindent where $Z = \sum_{J=0}^\infty (2J+1) \exp{(-hB_e J(J+1)/kT)}$
is the Partition function. Here, $B_e$ is the rotation constant. 

Assuming a fiducial CO interstellar abundance of $[{\rm
    H}_2]/[\doce]=10^4$ and $[\doce]/[\trece]=76$, the disk gas mass
inside the dust cavity as estimated from \trece\ is given by
\begin{equation}
  M_{\rm gas} = N(\rm total) \left [ \frac{\doce}{\trece} \right
  ]\left [ \frac{{\rm H}_2}{\doce} \right ] \mu m_{\rm H_2} \Omega \,
  d^2,
\end{equation}
\noindent where $m_{\rm H_2}$ is the molecular mass of hydrogen, $\mu
= 1.36$, $\Omega$ is the solid angle subtended by the cavity in
\trece, and $d =140$~pc is the distance to HD~142527.  For the $J=2-1$
CO isotopologues, $B_e = 55.1$~GHz and $A_{21} = 3.04 \times
10^{-7}~{\rm s}^{-1}$ for the line strength\footnote{The spectral
  information for the CO molecule was obtained from the Splatalogue
  database.}.  For the gas temperature we used $T=43$~K, as calculated
from \doce\ in the previous section.

We measured the integrated line intensity of \trece\ by summing the
peak emission from the selected high velocity channels (see
Section~\ref{sec:channels}), under the approximation that the
subtended solid angle is the same for all channels, and that the
source is unresolved at these velocities, yielding $\hat{I} = 4.9
\times 10^{-10}$~W~m$^{-2}$. This $\hat{I}$ gives a total column
density of $N = (8.69 \pm 0.7) \times 10^{20}$~m$^{-2}$. This column
corresponds to a total disk gas mass (H$_2$ plus CO) of $M_{\rm gas} =
(1.7 \pm 0.6)\times 10^{-3}$~M$_\odot$. Here, we have considered all
the gas inside the dust-depleted cavity of 140~au. The quoted thermal
errors are 5-$\sigma$ (based on the noise levels in Table~1) and are
subject to a further $\sim$10\% flux calibration systematic
uncertainty.  The same procedure, if applied to the
\dieciocho\ emission, yields a gas mass of $M_{\rm gas} \sim 2 \times
10^{-3}$~M$_\odot$.

Archival ALMA observations of \trece~$J=3-2$ also show strong emission
at channels with velocities $<+1.8$~km~s$^{-1}$ and
$>+5.4$~km~s$^{-1}$, corroborating our detection of CO $J=3-2$ gas
inside the cavity \citep[see][for a description of this
  dataset]{Fuk2013}. These observations show that \trece~$J=3-2$ is
mostly optically thick over the entire disk.  The noise level on each
of these channel map is 0.12~mJy~beam$^{-1}$ \citep{Fuk2013}. If we
assume \dieciocho~$J=3-2$ is optically thin at least inside the
cavity, the integrated line intensity amounts to $\hat{I} = 8.23
\times 10^{-10}$~W~m$^{-2}$, which gives a total column of $1.1 \times
10^{20}$~m$^{-2}$ and a disk mass of $(1.1 \pm 0.6)\times
10^{-3}$~M$_\odot$, consistent with the $J=2-1$ estimate (5-$\sigma$
error bars). For this calculation we used the same procedure explained
above to estimate the integrated line intensity arising from the
cavity gas but with the corresponding constants, $B_e = 54.9$~GHz and
$A_{32} = 2.17 \times 10^{-6}~{\rm s}^{-1}$, for the C$^{18}$O $J=3-2$
molecule.

We conclude that there is still at least more than one Jupiter mass
worth of gas inside the dust-depleted cavity. This gas is available to
be accreted by possible forming planets and/or to be accreted onto the
star. However, there are considerable uncertainties involved in this
analysis, making this result only a crude estimate of the total gas
mass. The main sources of uncertainty are the assumed isotopic line
ratios and the gas temperature.

\section{Physical Structure of the Disk}

\subsection{Parameterized Model}
\label{sec:model}
\begin{figure}
\epsscale{1.}
\plotone{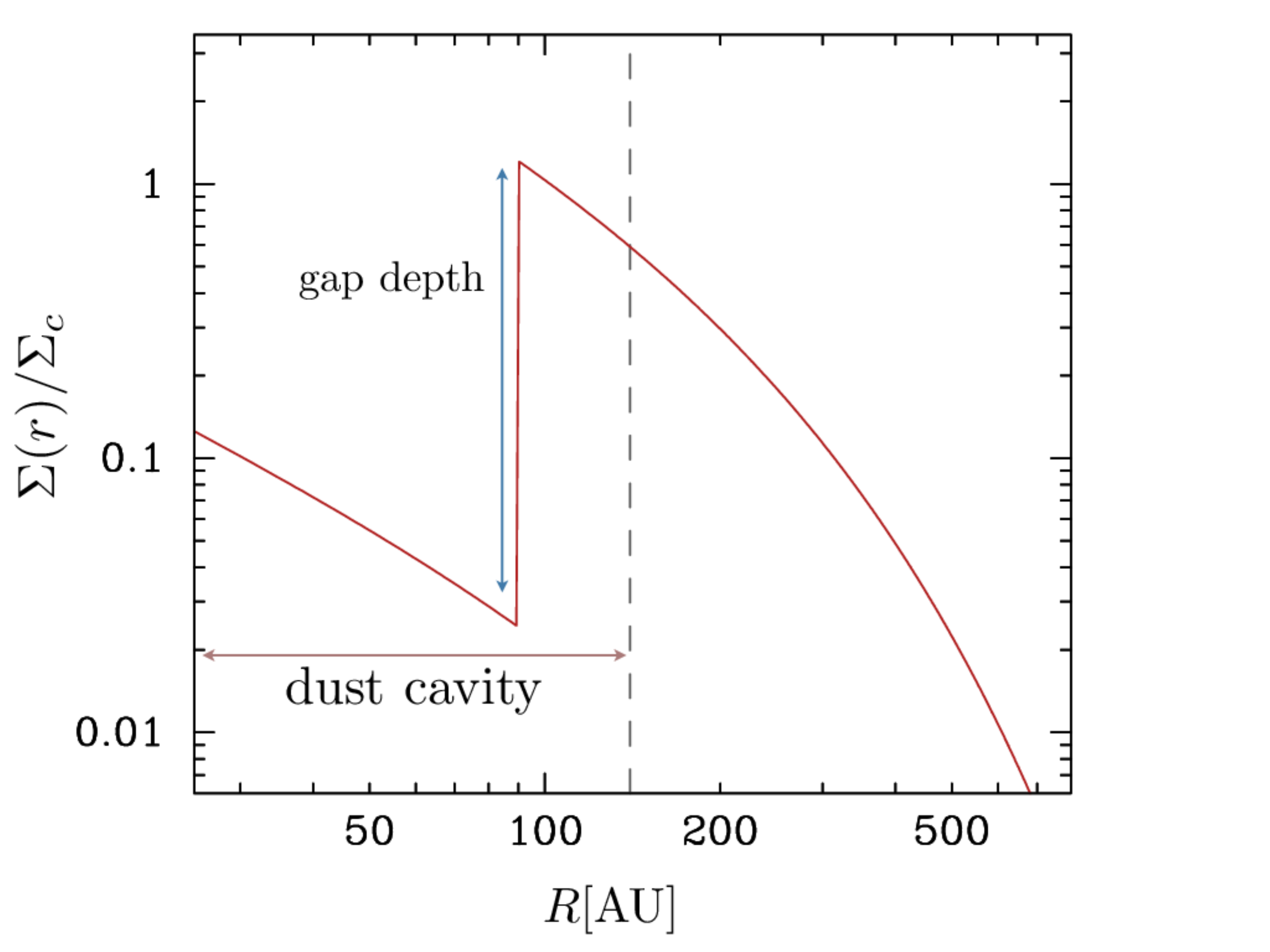}
\caption{Model surface density profile (Equation~\ref{eq:sigma}) of
  the disk. }.\label{fig:sigma}
\end{figure}

We adopted a simple, azimuthally symmetric parameterized structure for
the disk gas density, where the vertical dimension is given by the
hydrostatic equilibrium equation, such that,
\begin{equation}
\rho (r,z) = \frac{\Sigma(r)}{\sqrt{2\pi} h}
\exp{\left[-\frac{1}{2}\left( \frac{z}{h}\right)^{2}\right]},
\label{eq:rho}
\end{equation}
\noindent where $h(r)=0.1 r$ is the disk scale-height. The inner wall
of the outer disk needs to subtends a large solid angle in order to
explain the SED, as informed by \citet{Ver2011}. Eq.~\ref{eq:rho}
assumes the disk is isothermal in the vertical direction.  The global
surface density profile is a power law in radius with an exponential
tapering at large radii, as described in \citet{Lyn1974} \citep[see
  also][]{And2011}. The parametric equation reads
\begin{equation}
\Sigma (r) = \delta_{\rm gap}~\Sigma_c \left( \frac{r}{R_c}\right)^{-\gamma}
\exp{\left[-\left( \frac{r}{R_c}\right)^{2-\gamma}\right]},
\label{eq:sigma}
\end{equation}
\noindent where $R_c$ is a characteristic radius and $\Sigma_c$ is the
surface density at that radius. The model disk structure is then
separated into two distinct radial zones: a depleted inner cavity and
an outer disk. The $\delta_{\rm gap}$ parameter is a scaling factor
which accounts for the depletion inside the cavity ($r < R_{\rm
  cav}$). $\delta_{\rm}$ is a free parameter inside the cavity, and
attains unity in the outer disk. See Fig~\ref{fig:sigma} for a
schematic representation of the surface density model. The inner zone
of the disk is truncated at 0.3~au, at which the temperature reaches
$T=1500$~K, sublimating dust particles.

The kinetic temperature of the gas is flat close to the star
($r<50$~au), with a warm temperature of 50~K. Farther out, we modelled
the temperature with a power law, such that:
\begin{equation}
  T(r) = 50 \left(\frac{r}{50 {\rm au}} \right)^{-0.3} ~{\rm K},
\label{eq:temp}
\end{equation}
\noindent where the power law slope and scaling factor were informed
by the \doce\ optically thick emission (See
Section~\ref{sec:temp}). The disk is assumed to be vertically
isothermal.

For simplicity, we assumed that the gas in the disk follows circular
Keplerian rotation, such that

\begin{equation}
v_{\rm k}(r) =\sqrt{ \frac{G M_\star}{r}},
\label{eq:vel}
\end{equation}

\noindent where $M_\star$ is the central mass star.  Although hints of
non-Keplerianity are seen in the \doce\ channel maps, these have a
minor effect in our analysis.  The velocity dispersion of the random
motions of the gas (Doppler b-parameter) was fixed to 50 m~s$^{-1}$.

The CO column density is given by:
\begin{equation}
N_{\rm CO} (r) = \frac{X_{\rm CO}}{\mu m_{\rm H}} \Sigma (r),
\label{eq:nco}
\end{equation}
\noindent where $X_{\rm CO}$ is the abundance of CO relative to
H$_2$. We have assumed the ISM value.

Equations \ref{eq:rho}--\ref{eq:nco} fully describe the structure of
the disk. The disk is then characterized by six parameters,
$\Sigma_c$, $R_c$, $\gamma$, $R_{\rm cav}$, $M_\star$, and inclination
$i$. The minor axis position angle is $\sim$70 degrees as determined
from the velocity maps. In order to compare with the Band 6
observations, the synthetic models of line emission were resampled and
transformed to visibilities using the exact baseline information from
the ALMA measurement set.  We also calculate HD~142527's spectral
energy distribution (SED) for our disk structure model. The results
show consistency between our model and HD~142527's bulk properties
(see Appendix A).  It is important to note that these disk models have
degeneracies.  The models also assume an axisymmetric structure for
the gas, but from HCO$^+$ observations we know there is a filamentary
structure inside the cavity.

\begin{figure}
\epsscale{1.}
\plotone{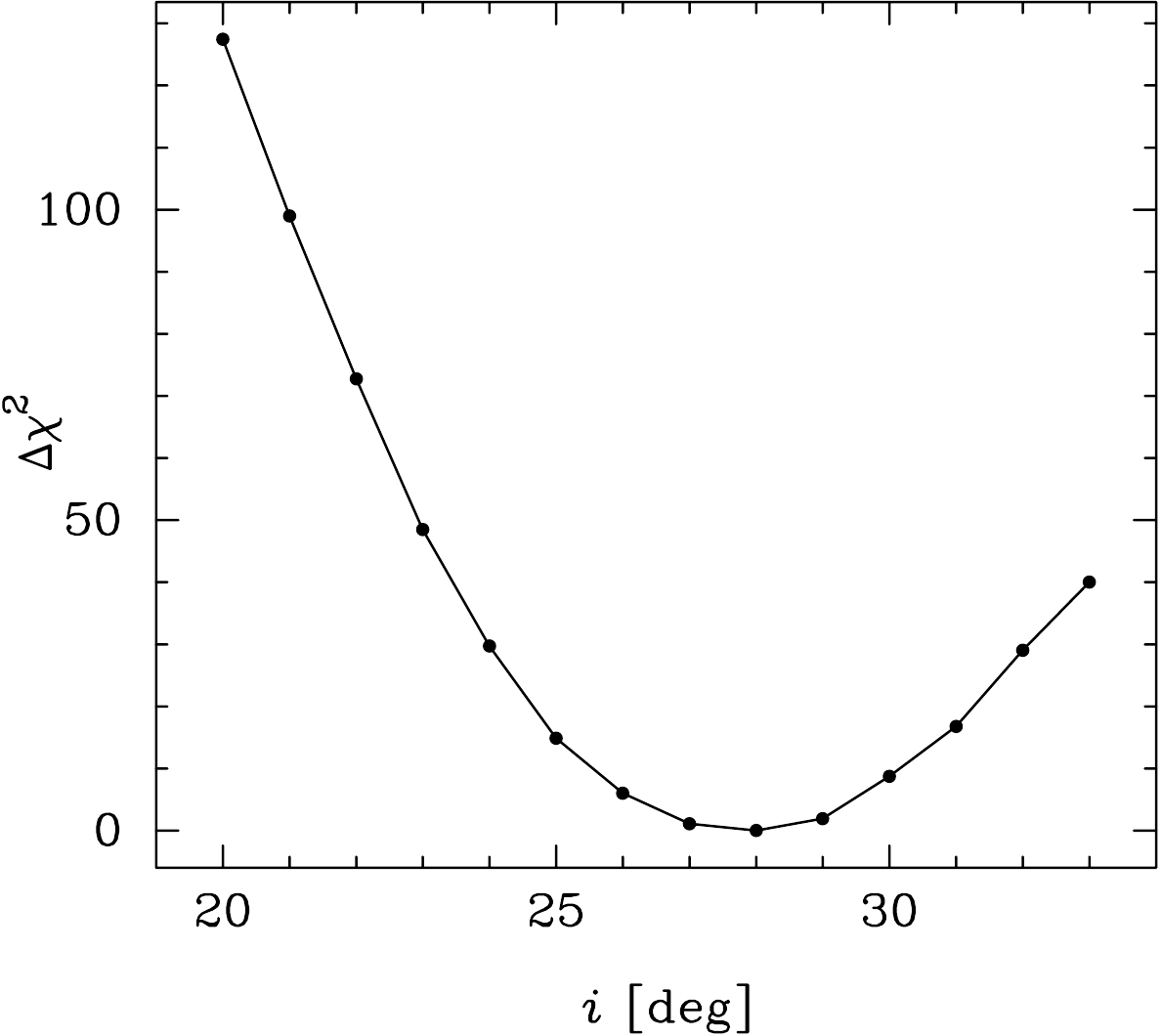}
\caption{Plot of $\Delta\chi^2$ as a function of a single parameter,
  the inclination of the disk $i$. The curve has a minimum around
  $i=28$~deg, with a 1-$\sigma$ error of $\pm$0.5~deg. }
\label{fig:ichi2}
\end{figure}

\subsection{Modeling Results}

\begin{deluxetable}{lccc}
\tabletypesize{\scriptsize}
\tablecaption{Parameters of gas disk structure}
\tablewidth{0pt}
\tablehead{
\colhead{Parameter} & Units & \colhead{Value}  & \colhead{$1\sigma$ error} }
\startdata
$i$ & deg & 28 & 0.5\dag\\
$R_{\rm cav}$ & au & 90 & 5\\
$\delta_{\rm gap}$ &  & 0.02 & \\
$M_{\rm cav}$ & M$_\odot$ & $2\times 10^{-3}$ &  $1\times 10^{-3}$\\ 
$\log \Sigma_c$ & ${\rm m}^{-2}$& 28.69 & --\\
$M_{\star}$ & M$_\odot$ & 2.2 & --\\ 
$h/r$ & -- & 0.1 & -- \\
$\gamma$ & -- & 1 & -- \\
$R_c$ & au & 200 & -- \\
$V_{\rm sys}$ & \kms & 3.6 & 0.1
\enddata
\tablecomments{We report 1~$\sigma$ error bars. Only parameters that
  have been optimized bear uncertainties. \dag Assuming a fixed star mass
  of 2.2 M$_\odot$.}
\label{tab:params}
\end{deluxetable}

We predicted the line radiation from the set of model disks using the
three-dimensional non-local thermodynamic equilibrium (non-LTE)
molecular excitation and radiative transfer code LIME \citep{Bri2010}.
We generated channel maps matching the spectral resolution of the ALMA
data cubes. Then, we extracted the $uv$-coverage information directly
from the ALMA measurement sets to calculate model visibilities that
match the exact Fourier sampling used in the observations.  Our models
were compared to the data in the visibility plane based on $\chi^2$
statistics.  The models were post-processed and CLEANed using the same
strategy used to generate our ALMA maps. Finally, the best fit models
were contrasted with the data by visual inspection.

\begin{figure*}
  \centering\includegraphics[width=0.75\textwidth]{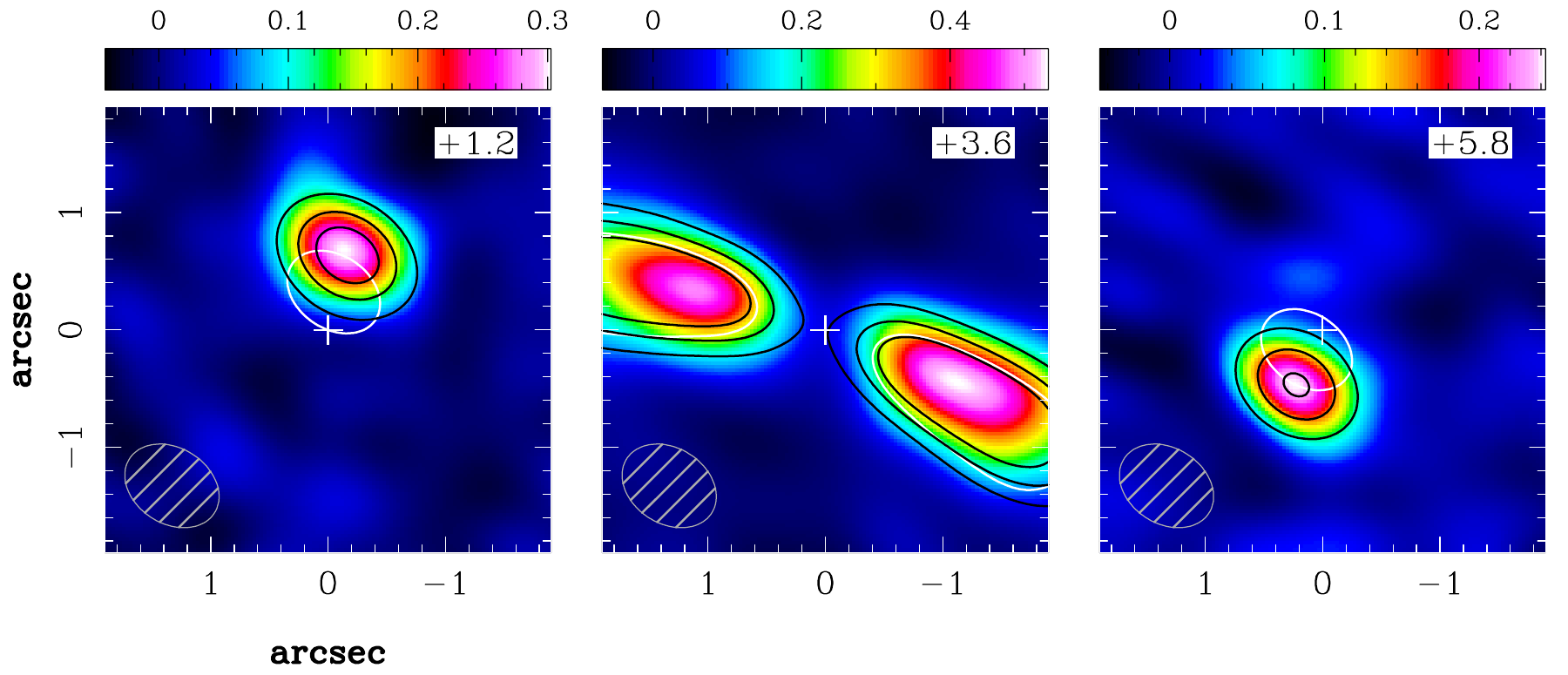}
  \caption{These panels show the bestfit model for the disk structure
    in HD~142527 in \trece\ emission. The model is presented in
    countours while the ALMA data are shown in false color. Black
    countours show the model with an inclination angle of 28 degrees
    and a central star mass of 2.2~M$_\odot$, which best reproduces
    the observed morphology. The white countours show a similar model
    but with a disk inclination of 20 degrees only. {\it Left:} Blue
    high velocity channel that shows gas coming from inside the
    cavity, close to the central. {\it Center:} Systemic velocity
    channel. {\it Right:} Red high velocity channel. See top wedge for
    contours and emission levels. }
\label{fig:model}
\end{figure*}

Figure~\ref{fig:model} shows our best fit model for the structure of
HD~142527 transition disk that accounts for the observed
\trece\ emission. \trece\ is a tracer of the disk density
distribution, it is optically thin inside the cavity and the maps have
high signal-to-noise ratio.  The best fit model also matches the
\dieciocho\ maps. The integrated density distribution of our disk
model yields a total (dust plus gas) mass of the disk of
0.1~M$_{\odot}$, consistent with previous estimates based on
millimeter continuum observations \citep{Obe2011}.

The model parameters that match the observations are listed in
Table~\ref{tab:params}.  A tapered disk slope of $\gamma = 1$ accounts
for the extended size of the source in \doce.  Inside the cavity, The
$\delta_{\rm gap}$ parameter was set to $0.02$, in order to reproduce
our observed line fluxes. Higher values make the \trece\ appear
optically thick inside the cavity. We estimate an uncertainty of 0.01
based on the optical thickness of the lines. It is worth mentioning
that the slope of the density profile inside the cavity does not have
to be necessarily decreasing with radius \citep[see, for example,][for
  HD100546]{Tat2011}. A flat profile inside the cavity might also be
valid, which would imply a shallower gap depth value. These
posibilities will be further explored in a future paper.

Our method to estimate the inclination angle differs from previous
approaches because it relies on the dynamics of the high velocity gas
evidenced by the high velocity channels, whose spatial distribution
strongly depends on inclination angle. Previous methods use scattered
light and continuum images of the full disk. Continuum images trace
the emitting material closer to the mid-plane hence giving an estimate
of the inclination of the dust near the midplane.  Our approach uses
the dynamical model of the gas rotation, giving a good estimate of the
gaseous disk's inclination. 

There is a degeneracy between inclination angle, $i$, and the mass of
the central star, $M_\star$, when trying to compute the kinematic
signatures of the rotating gaseous disk. In order to break this
degeneracy, we run a set of test models for a range of values of
$M_\star$ and $i$.  The goodness of each fit was evaluated by
calculating the reduced $\chi^2$, which is defined as $\chi_{\nu}^2 =
\chi^2/\nu$, where $\nu$ is the number of the degrees of freedom. For
these observations, the number of degrees of freedom is 10$^7$, which
corresponds to the number of visibilities in each dataset.  The
minimum $\chi_\nu^2$ (best fit) is attained at $M_{\star} =
2.2$~M$_\odot$ and it is equal to 1.001662.  On the other hand, the
values of $\Delta \chi_\nu^2$ (defined as $\chi_\nu^2 - {\rm
  min}\{\chi_\nu^2\}$) for $M_\star$ equal to 2 and to 2.6~M$_\odot$
are 2$\times$10$^{-6}$ and 5$\times$10$^{-6}$, respectively.  Similar
values were found for inclination angles ranging from 20 to 30
degrees. According to our $\chi_\nu^2$ fit in the visibility plane,
the model with a central star mass $M_\star=2.2$~M$_{\odot}$ best
reproduces our \trece\ data.  After fixing the mass to this value, we
estimate the disk's inclination angle.

Previous models for the disk in HD~142527 assumed a cavity size of
140~au for both gas and dust and a disk inclination of 20 degrees
\citep{Ver2011, Cas2013}. Although the overall multi-wavelength data
are consistent with these parameters, they are unable to explain the
observed spatio-kinematic morphologies of the CO emission lines. In
order to reproduce the high velocity channel maps (wings of the
lines), a significant increase in the projected velocity closer to the
central star is needed.  With a large disk inclination, $i= 28 \pm
0.5$ degrees, we obtain a good match with the ALMA data, reproducing
the \trece\ morphologies both at systemic velocity and in the high
velocity wings. The excellent signal-to-noise in the high velocity
channel maps allows us to determine the inclination of the disk with
an accuracy of half a degree. Figure~\ref{fig:ichi2} shows a $\chi^2$
plot for a distribution of inclination angles, calculated in the
$uv$-plane.

After fixing the inclination angle at 28 deg, we varied the cavity
radius in order to find a good match.  The best fit for the cavity
radius in the gas is 90$\pm5$~au. The error was obtained from visual
inspection of the plotted models against the \trece\ data.  It is
important to note that the inclination angle best characterises the
emission coming from inside the cavity.  Our estimate of the
inclination angle is consistent with the values determined by
\citet{Fuk2013} from CO $J=3-2$ observations.  

\section{Discussion}\label{sec:discussion}

We have conducted a detailed analysis of the \doce, \trece, and
\dieciocho\ $J=2-1$ line emission from HD~142527, focusing on the gas
inside the dust-depleted cavity.

There are three main scenarios for clearing central cavities in disks;
photoevaporation by high-energy photons, lowering of dust opacity due
to grain-growth and dynamical clearing by a companion
object. \citet{Cas2013} argued against photoevaporation in this
source, based mainly on the high accretion rate and large size of the
cavity.  Our detection of large amounts of CO gas inside the cavity
furthers the argument against photoevaporation as the main clearing
mechanism. Moreover, there is evidence of the presence of a small
inner disk inside the cavity \citep{vBoe2004, Ver2011}, that would
hinder the efficiency of photoevaporating winds.

Dust continuum emission at 345~GHz has been detected with ALMA inside
the cavity, but at very low levels and at the limit of the dynamic
range.  The exact morphology of this faint signal is sensitive on the
details of image synthesis and self-calibration process. While
\citet{Cas2013} find a filamentary morphology, matched to the
gap-crossing filaments seen in HCO$^+$(4-3), \citet{Fuk2013} proposes
a compact signal stemming from the inner disk only. Whichever the
case, the peak signal at 345~GHz is an upper limit to the emission
from mm-sized particles inside the cavity. The contrast between the
bottom of the gap and the outer disk in surface density of mm-emitting
grains is $<300$ (gap depth of about 0.003\%).

Smaller dust particles, $\sim 1~\mu$m dust emitting in the near and
mid-IR, have eluded detection inside the 90~au gas cavity radius
\citep{Fuk2006, Cas2012, Can2013}, although the IR rings are contained
in the sub-millimeter ring \citep[see Supplementary information
  in][]{Cas2013}. This may imply that all the dust follows a general
drop in density inside the 90~au cavity.  The polarimetric data
provides upper limits on the surface brightness of the polarized
intensity emitted from inside the cavity \citep{Can2013}.

The current paradigm in giant planet formation holds that planets are
manifested indirectly by their imprinted marks on their progenitor
disk.  Wide gaps will be carved by massive or multiple forming planets
\citep{Dod2011, Zhu2011}.  The shape of the outer disk inner wall will
depend on the planet mass, disk thickness and viscosity of the disk,
and on the temperature structure in the disk wall itself
\citep{Cri2006}.  The torque of the outermost planet will affect
strongly the shape of the inner wall. The tidal radius of a planet,
the so-called Hill radius, is defined by
\begin{equation}
R_{\rm H} = a_p \left ( \frac{M_p}{3M_\star} \right )^{1/3},
 \label{eq:hill}
\end{equation}
\noindent where $a_p$ is the radius of the planet's orbit, $M_p$ is
the mass of the planet, and $M_\star$ is the mass of the central
star. More massive companions have a large sphere of influence,
suggesting that they could carve larger gaps with attenuated or
tapered-off walls.

The cavity in HD~142527 has a smaller radius in gas than in dust
milimeter continuum. In HCO$^+$(4-3) emission the cavity is even
smaller, suggesting that the disk wall is not sharp or vertical but
rather tapered off by a large gradient (see Fig.~\ref{fig:cut}).
Multiple planets located closer to the star can explain the large size
of the cavity \citep{Dod2011, Zhu2011}. A precise analysis of the gas
and dust wall in HD~142527 from ALMA band 6 and band 7 data, based on
hydrodynamical models, will be presented in a future publication.

\section{Summary} \label{sec:summary}

We detected and identified emission from inside the dust-depleted
cavity in carbon monoxide isotopologue data of the gas-rich transition
disk \objectname{HD~142527}, in the $J=2-1$ line of \doce,
\trece\ and \dieciocho. The $^{12}$CO emission from the inner cavity
is optically thick, while \trece\ and \dieciocho\ appear to be
optically thin, providing probes of the temperature and density fields
inside the cavity.  The main results of our analysis can be summarized
as follows.

\begin{enumerate}
\item We determined the gas mass inside the gap from the $^{13}$CO and
  C$^{18}$O emission. We used two methods: direct measurement of the
  optically thin lines, and forward modelling using an axysymmetric
  disk. The total mass of gas surviving inside the cavity is high
  $(1.7\pm 0.6) \times 10^{-3}$~M$_\odot$.
\item We find that the inner cavity is rather small in CO gas,
  compared to its size in dust millimeter and infrared scattered
  light, with a best-fit radius of 90$\pm 5$~au. 
\item The drop in density inside the cavity can be modeled as a
  reduction in gas column density of a factor 50 (cavity depth of
  0.02), given our model for the outer disk, which gives a total mass
  of 0.1~M$_\odot$.
\item The gap wall appears diffuse and tapered-off in the gas
  distribution, while in dust continuum is manifestly sharper.
\item The disk inclination is well constrained by the resolved
  velocity information of the CO gas and it attains 28$\pm$0.5
  degrees, after fixing the central star mass to 2.2~M$_\odot$.
\item The center of the disk appears shifted from a central continuum
  emission by a deprojected distance of $\sim$16~au toward the East,
  assuming a distance of 140~pc to HD~142517.
\end{enumerate}

\acknowledgments

 We thank the anonymous referee for helpful comments and
 suggestions. SP, GvdP, SC and FM acknowledge support from the
 Millennium Science Initiative (Chilean Ministry of Economy), through
 grant ``Nucleus P10-022-F''.  SP, GvdP, SC, and PR acknowledge
 financial support provided by FONDECYT following grants 3140601,
 3140393, 1130949, and 3140634.  This paper makes use of the following
 ALMA data: ADS/ JAO.ALMA\#2011.0.00465.S. ALMA is a partnership of
 the ESO, NSF, NINS, NRC, NSC and ASIAA. The Joint ALMA Observatory is
 operated by the ESO, AUI/NRAO and NAOJ. The National Radio Astronomy
 Observatory is a facility of the National Science Foundation operated
 under cooperative agreement by Associated Universities, Inc.
 P.R. acknowledges support from Project ALMA-CONICYT
 \#31120006. C.P. acknowledges funding from the European Commission's
 7$^\mathrm{th}$ Framework Program (contract PERG06-GA-2009-256513)
 and from Agence Nationale pour la Recherche (ANR) of France under
 contract ANR-2010-JCJC-0504-01. L.C. acknowledges support from
 Project ALMA-CONICYT \#31120009.

\appendix

\section{Spectral energy distribution}\label{sec:sed}

We tested the consistency of our disk structure model by comparing its
predicted spectral energy distribution (SED) against observed
HD~142527 photometry.  We obtained observed continuum fluxes from
\citet{Ver2011} (see references there in).  The observed SED was
dereddened by using a standard extinction curve \citep{Car1989}, for
$A_{\rm V}=0.6$ \citep{Ver2011} and an assumed $R_{\rm V}$ of $3.1$.

Synthetic fluxes were calculated using the Monte Carlo radiative
transfer code MCFOST \citep{Pin2006}.  We used an MCFOST model
consistent with the disk structure described in
Section~\ref{sec:model}.  Previous attempts to model the HD~142527 SED
invoked a puffed-up rim to explain the large excess of emission in the
near infrared \citep{Ver2011}. This excess could be explained by the
dust emission coming from very near the star that we see at 345~GHz
(see Section~\ref{sec:cut}).

We distributed dust grains (mainly silicates and amorphous carbons)
with sizes between 0.1 and 1000~$\mu$m across the disk model. We used
the Mie theory and assumed a gas-to-dust ratio of 100.

\begin{figure}
\epsscale{1.}
\centering
\plotone{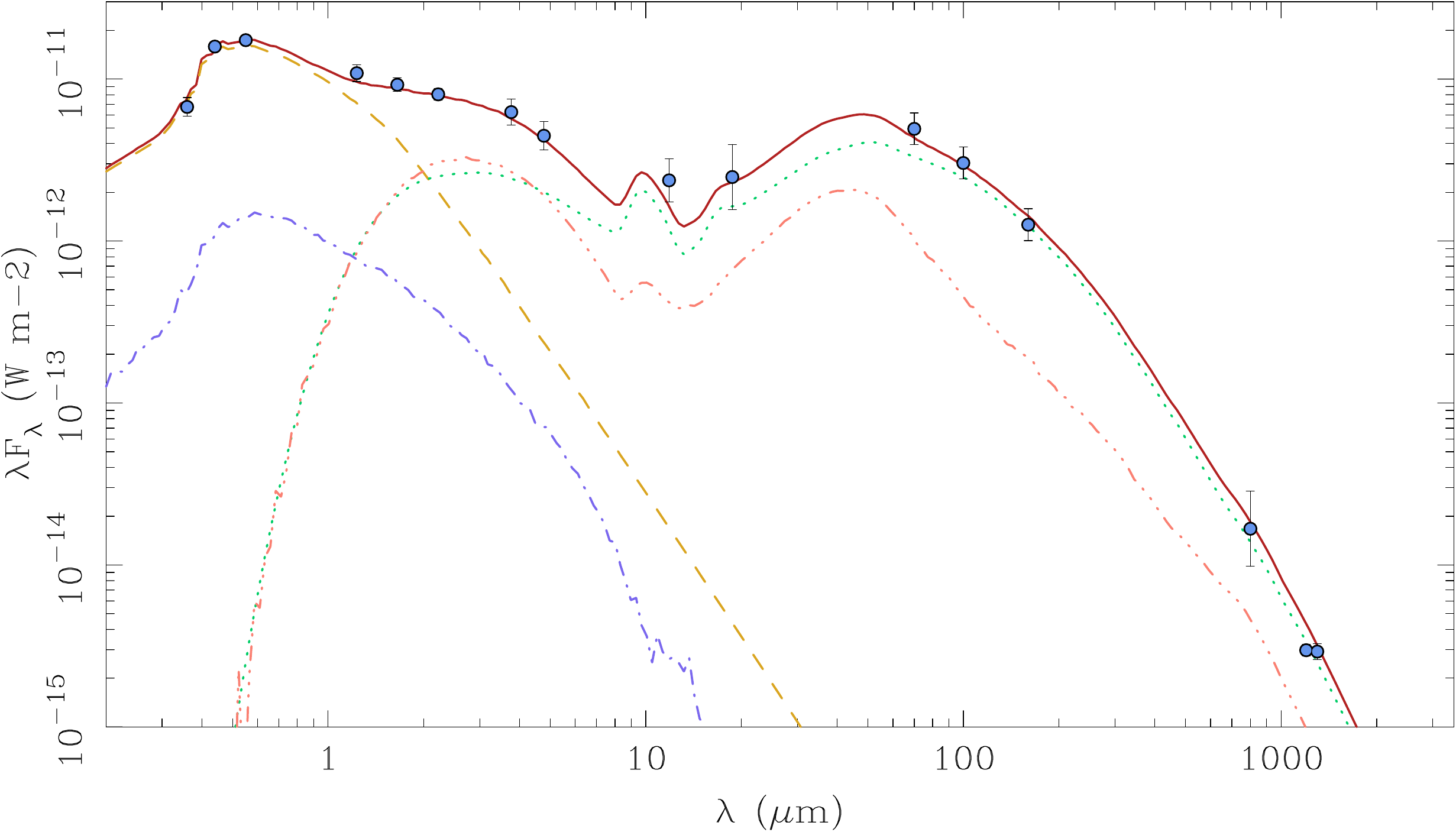}
\caption{Dereddened photometry (blue dots) and spectral energy
  distribution model of HD142527. Photometric data points have been
  obtained from \citet{Ver2011}. We have added the new ALMA continuum
  photometry at 1.2~mm. 5$\sigma$ error bars are shown. The model
  contributions are: stellar spectrum alone (dashed yellow curve),
  scattered light from the star (dash-dotted light blue curve),
  thermal emission from inner and outer disks (dotted green curve),
  scattered light from the dust thermal emission (dash triple dotted
  orange curve). The resulting model is the red solid curve.}
\label{fig:sed}
\end{figure}

Figure~\ref{fig:sed} shows the SED calculated from our model.  Our
model consists of: 1) an inner disk extending from 0.2 to 6~au with
10$^{-9}$~M$_\odot$ of dust. This dust mass, together with a aspect
ratio $h/r$ of 0.18, are enough to reproduce the large near-infrared
excess. 2) a dust-depleted gap between 6 and 90~au in radii almost
depleted of gas and dust; 3) a second gap extending from 90 to 130~au
with enough gas to explain our isotopologues observations, and 4) a
large outer disk stretching from 130 to 300~au.  The flaring exponent
ranges between 1 and 1.18 for each zone. The aspect ratio of all the
zones was 0.18 except for the outer disk which covers a larger solid
angle, attaining an $h/r$ of 0.23.  

It is important to note that the gas model presented in
Section~\ref{sec:model} assumes a constant aspect ratio, since it only
aims to reproduce the gas inside the dust-depleted cavity and not the
broad spectral features of the entire disk.

\end{document}